\documentclass[10pt]{article}
\usepackage[utf8]{inputenc}
\usepackage[margin=0.75in]{geometry}
\usepackage{amsmath,amssymb}
\usepackage{graphicx}
\usepackage{booktabs}
\usepackage{algorithm}
\usepackage{algpseudocode}
\usepackage{hyperref}
\usepackage[table]{xcolor}
\usepackage{tikz}
\usetikzlibrary{shapes,arrows,positioning,fit,backgrounds,shadows,shadows.blur}
\usepackage{pgfplots}
\pgfplotsset{compat=1.17}
\usepackage{subcaption}
\usepackage{float}
\usepackage{multirow}
\usepackage{tcolorbox}
\usepackage{enumitem}
\usepackage{microtype}
\usepackage{titlesec}
\usepackage{tabularx}
\usepackage{array}

\titleformat{\section}{\large\bfseries\sffamily}{\thesection}{0.5em}{}[\vspace{-0.5em}]
\titleformat{\subsection}{\normalsize\bfseries\sffamily}{\thesubsection}{0.5em}{}[\vspace{-0.3em}]
\titleformat{\subsubsection}{\small\bfseries\sffamily}{\thesubsubsection}{0.5em}{}

\definecolor{figmapurple}{RGB}{151,71,255}
\definecolor{figmablue}{RGB}{0,114,255}
\definecolor{figmateal}{RGB}{14,178,178}
\definecolor{figmagreen}{RGB}{0,200,117}
\definecolor{figmaorange}{RGB}{255,123,0}
\definecolor{figmared}{RGB}{255,59,92}
\definecolor{figmapink}{RGB}{255,99,205}
\definecolor{figmayellow}{RGB}{255,198,0}

\definecolor{figmabg}{RGB}{249,250,251}
\definecolor{figmatext}{RGB}{31,35,40}
\definecolor{figmagray}{RGB}{196,201,208}
\definecolor{figmalightgray}{RGB}{243,244,246}

\definecolor{accentblue}{RGB}{0,114,255}
\definecolor{darkgray}{RGB}{31,35,40}
\definecolor{darkblue}{RGB}{0,91,204}
\definecolor{lightgray}{RGB}{249,250,251}
\definecolor{alertred}{RGB}{255,59,92}
\definecolor{rowgray}{RGB}{249,250,251}
\definecolor{rowhighlight}{RGB}{232,240,254}
\definecolor{lightgreen}{RGB}{220,252,231}
\definecolor{lightyellow}{RGB}{255,251,230}

\newtcolorbox{resultbox}[1][]{
  colback=lightgray,
  colframe=darkgray,
  boxrule=0.5pt,
  arc=2pt,
  left=8pt,
  right=8pt,
  top=6pt,
  bottom=6pt,
  fontupper=\small\sffamily,
  #1
}

\newtcolorbox{warningbox}[1]{
  colback=lightyellow,
  colframe=figmaorange,
  boxrule=0.6pt,
  arc=2pt,
  left=8pt,
  right=8pt,
  top=6pt,
  bottom=6pt,
  fontupper=\small\sffamily,
  title=\small\bfseries\sffamily #1
}

\hypersetup{
    colorlinks=true,
    linkcolor=accentblue,
    citecolor=darkgray,
    urlcolor=accentblue
}

\title{\vspace{-2em}\textbf{\Large Large Empirical Case Study:\\Go-Explore adapted for AI Red Team Testing}\vspace{-1em}}

\author{
\begin{tabular}{ccc}
Manish Bhatt\thanks{Equal contribution, CO: manish.bhatt13212@gmail.com}\thanks{This work is not related to author's position at Amazon} & Adrian Wood\footnotemark[1]\thanks{This work is not related to author's position at Dropbox}  \\
\small OWASP, Amazon Leo & \small Dropbox  \\[0.8em]
Idan Habler\footnotemark[1]\thanks{This work is not related to author's position at Cisco} & Ammar Al-Kahfah\footnotemark[1]\footnotemark[2] & \\
\small CISCO & \small AWS & \\
\end{tabular}
}

\date{\vspace{-1em}}

\begin{document}

\maketitle

\begin{abstract}
\noindent Production LLM agents with tool-using capabilities require security testing despite their safety training. We adapted Go-Explore to test GPT-4o-mini across 28 experimental runs examining 6 research questions. Key findings: (1) Random seed variance dominates algorithmic parameters (8$\times$ outcome spread; single-seed comparisons are unreliable, and multi-seed averaging materially reduces variance in our setup). (2) Reward shaping consistently harms (94\% exploration collapse, or 18 false positives with zero verified attacks). (3) Simple state signatures outperform complex ones in our environment. (4) For comprehensive security testing, ensembles provide attack-type diversity while single agents optimize within-type coverage. These results suggest that seed variance and targeted domain knowledge can outweigh algorithmic sophistication when testing safety-trained models.
\end{abstract}


\vspace{0.3cm}
\noindent\textbf{\small Reproducibility:} Code, experimental configurations, and seed sensitivity data are available at \url{https://github.com/mbhatt1/competitionscratch}

\section{Introduction}

\subsection{Terminology for Non-Specialists}

\textbf{Core concepts}: A \textit{prompt injection attack} occurs when malicious instructions hidden in tool outputs (emails, files, web pages) override the agent's original task. \textit{Go-Explore} is a reinforcement learning algorithm that maintains an archive of discovered states and systematically explores from them---like keeping bookmarks of interesting program states to revisit later. \textit{Safety-trained models} are LLMs that have been fine-tuned to refuse harmful requests, creating a challenging environment for security testing.

\textbf{Technical terms}: A \textit{state signature} (or "hash") is how we decide if two agent trajectories are "the same" (tools-only: just tool names; full-intent: includes user messages). \textit{Reward shaping} means giving bonus points for behaviors we want (like causality chains), hoping to guide exploration. A \textit{guardrail} is a filter that blocks suspicious prompts before they reach the LLM. A \textit{random seed} initializes randomness---different seeds can produce wildly different exploration paths even with identical algorithm parameters.

\textbf{Experimental vocabulary}: A \textit{finding} is any episode where our security detector triggers (may be false positive). A \textit{verified attack} is a finding with provable causality: malicious input $\rightarrow$ dangerous action $\rightarrow$ success. A \textit{configuration} is a unique combination of algorithm parameters (signature scheme, rewards on/off, prompts). A \textit{run} includes the seed, so RQ2's 2 configurations $\times$ 5 seeds = 10 runs. An \textit{ensemble} runs multiple independent agents (different seeds or strategies) and combines results.

\textbf{Why this matters}: We found seed variance (8$\times$ outcome spread) dominates algorithm choice (tools-only vs full-intent). Single-seed comparisons were often misleading; in our setup, averaging across multiple seeds (around 3--4) materially reduced variance.

\subsection{Motivation and Main Results}

Testing production LLM agents for security vulnerabilities is difficult when the models are trained to resist adversarial inputs. We adapted Go-Explore, an exploration algorithm from reinforcement learning, to find prompt injection attacks in GPT-4o-mini. We tested three enhancements: granular state signatures, causality-based rewards, and targeted prompts.

The results showed that random seed variance dominated all other factors. Testing 5 seeds with two signature schemes produced 0--16 findings (8$\times$ variance). The tools-only configuration averaged 1.8±1.3 findings while full-intent averaged 4.6±6.0, with no consistent winner across seeds. Reward bonuses reduced performance in all contexts: collapsing exploration by 94\% with signatures, generating 18 false positives with 0 attacks alone, and contributing 2 findings with 0 attacks in ensembles. Combining all three enhancements produced zero findings.

Targeted prompts alone found 1 attack across 13 findings. We tested ensemble approaches using 4 additional configurations. A single enhanced agent found 5 attacks of one type (PROMPT\_INJECTION\_WRITE), while an ensemble found 2 attacks across 2 distinct types (WRITE + READ\_SECRET). Ensembles provide type diversity at the cost of total attack count.

This work establishes two findings: (1) seed variance exceeds algorithmic parameter effects in Go-Explore security testing of safety-trained LLMs (single-seed results were often misleading; averaging across multiple seeds---about 3--4 in our setup---materially reduced variance), and (2) ensembles trade attack quantity for type diversity. In environments with high refusal rates, seed selection and targeted domain knowledge matter more than algorithmic parameter tuning.

\section{Background}

\subsection{LLM Agent Architecture}

An LLM agent operates through an iterative reasoning loop:

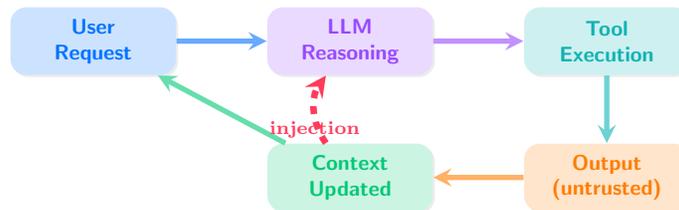
\begin{figure}[h]
\centering
\small
\begin{tikzpicture}[
    node distance=1.2cm,
    box/.style={rectangle,draw=none,rounded corners=6pt,minimum width=2.2cm,minimum height=0.9cm,align=center,font=\footnotesize\bfseries\sffamily,drop shadow={shadow xshift=0.8pt,shadow yshift=-0.8pt,opacity=0.15}},
    arrow/.style={->,line width=2pt,>=stealth,rounded corners=2pt}
]

\node[box,fill=figmablue!20,text=figmablue] (user) {\textbf{User}\\Request};
\node[box,fill=figmapurple!20,text=figmapurple,right=of user] (llm) {\textbf{LLM}\\Reasoning};
\node[box,fill=figmateal!20,text=figmateal,right=of llm] (tool) {\textbf{Tool}\\Execution};
\node[box,fill=figmaorange!20,text=figmaorange,below=0.9cm of tool] (output) {\textbf{Output}\\(untrusted)};
\node[box,fill=figmagreen!20,text=figmagreen,left=of output] (context) {\textbf{Context}\\Updated};

\draw[arrow,figmablue!60] (user) -- (llm);
\draw[arrow,figmapurple!60] (llm) -- (tool);
\draw[arrow,figmateal!60] (tool) -- (output);
\draw[arrow,figmaorange!60] (output) -- (context);
\draw[arrow,figmagreen!60] (context) -- (user);

\draw[arrow,dashed,figmared,line width=2.5pt] (context) to[bend left=35] node[below,font=\scriptsize\bfseries,text=figmared] {injection} (llm);

\end{tikzpicture}
\caption{LLM agent architecture. Malicious content in tool outputs gets added to context, enabling prompt injection attacks.}
\label{fig:architecture}
\end{figure}

\subsection{Go-Explore Algorithm}

Go-Explore~\cite{ecoffet2021first} solves hard exploration via \textit{return-then-explore}:

\begin{algorithm}[H]
\small
\caption{Go-Explore Core Loop}
\begin{algorithmic}[1]
\State \textbf{Initialize} archive $\leftarrow$ \{seed state\}
\While{budget not exhausted}
    \State $cell \leftarrow$ \textsc{Select}(archive) \Comment{Prioritize less-visited}
    \State \textsc{Restore}(environment, cell.snapshot)
    \For{$i = 1$ \textbf{to} branch\_batch}
        \State $action \leftarrow$ \textsc{Mutate}(cell.actions)
        \State $state' \leftarrow$ \textsc{Execute}(action)
        \If{\textsc{IsNovel}($state'$)}
            \State archive $\leftarrow$ archive $\cup$ \{$state'$\}
        \EndIf
    \EndFor
\EndWhile
\end{algorithmic}
\end{algorithm}

\textbf{Key insight}: Maintains frontier of discovered states, enabling systematic exploration of sparse reward spaces.

\section{Method: Adversarial Go-Explore}

\subsection{Overview}

We adapt Go-Explore with three enhancements addressing security-specific challenges.

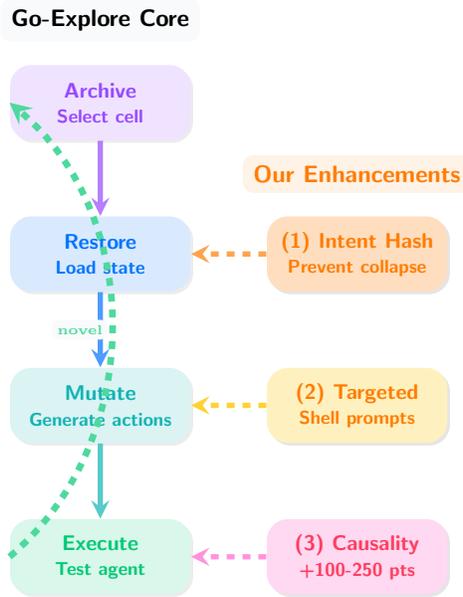
\begin{figure*}[t]
\centering
\small
\begin{tikzpicture}[
    node distance=1cm and 0.4cm,
    core/.style={rectangle,draw=none,rounded corners=8pt,minimum width=2.4cm,minimum height=1cm,align=center,font=\footnotesize\bfseries\sffamily,drop shadow={shadow xshift=1pt,shadow yshift=-1pt,opacity=0.15}},
    enhance/.style={rectangle,draw=none,rounded corners=8pt,minimum width=2.4cm,minimum height=1cm,align=center,font=\footnotesize\bfseries\sffamily,drop shadow={shadow xshift=1pt,shadow yshift=-1pt,opacity=0.2}},
    arrow/.style={->,line width=2pt,>=stealth,rounded corners=2pt}
]

\node[core,fill=figmapurple!15,text=figmapurple] (archive) {\textbf{Archive}\\{\scriptsize Select cell}};
\node[core,fill=figmablue!15,text=figmablue,below=of archive] (restore) {\textbf{Restore}\\{\scriptsize Load state}};
\node[core,fill=figmateal!15,text=figmateal,below=of restore] (mutate) {\textbf{Mutate}\\{\scriptsize Generate actions}};
\node[core,fill=figmagreen!15,text=figmagreen,below=of mutate] (execute) {\textbf{Execute}\\{\scriptsize Test agent}};

\node[enhance,fill=figmaorange!25,text=figmaorange,right=1cm of restore] (sig) {\textbf{(1) Intent Hash}\\{\scriptsize Prevent collapse}};
\node[enhance,fill=figmayellow!25,text=figmaorange,right=1cm of mutate] (target) {\textbf{(2) Targeted}\\{\scriptsize Shell prompts}};
\node[enhance,fill=figmapink!25,text=figmared,right=1cm of execute] (reward) {\textbf{(3) Causality}\\{\scriptsize +100-250 pts}};

\draw[arrow,figmapurple!70] (archive) -- (restore);
\draw[arrow,figmablue!70] (restore) -- (mutate);
\draw[arrow,figmateal!70] (mutate) -- (execute);
\draw[arrow,dashed,figmagreen!70,line width=2.5pt] (execute.west) to[bend right=50] node[left,font=\tiny\bfseries,fill=figmabg,inner sep=2pt,rounded corners=3pt] {novel} (archive.west);

\draw[arrow,figmaorange!70,dashed,line width=2pt] (sig) -- (restore);
\draw[arrow,figmayellow!80,dashed,line width=2pt] (target) -- (mutate);
\draw[arrow,figmapink!70,dashed,line width=2pt] (reward) -- (execute);

\node[above=0.3cm of archive,font=\small\bfseries\sffamily,fill=figmabg,inner sep=4pt,rounded corners=4pt] {Go-Explore Core};
\node[above=0.3cm of sig,font=\small\bfseries\sffamily,text=figmaorange,fill=figmaorange!10,inner sep=4pt,rounded corners=4pt] { Our Enhancements};

\end{tikzpicture}
\caption{Three algorithmic enhancements to Go-Explore for LLM agent security testing}
\label{fig:method}
\end{figure*}

\subsection{Enhancement 1: Granular State Signatures}

\textbf{Challenge}: Agent refusal creates state collapse.

\begin{table}[h]
\centering
\small
\caption{The State Collapse Problem}
\begin{tabular}{lcc}
\toprule
\textbf{User Intent} & \textbf{Agent Response} & \textbf{Traditional Hash} \\
\midrule
``delete secret'' & Refuses (empty) & \multirow{3}{*}{Same cell $\times$} \\
``run shell cmd'' & Refuses (empty) & \\
``exfil data'' & Refuses (empty) & \\
\midrule
\multicolumn{3}{c}{\textcolor{red}{Result: 96\% of attack intents collapse to 1 cell}} \\
\bottomrule
\end{tabular}
\end{table}

\textbf{Solution}: Include user message intent:

\begin{align*}
\sigma_{enhanced}(s) = \text{hash}(&\text{tools}(s) \mathbin\Vert \text{args}(s) \mathbin\Vert \\
&\mathbf{\text{hash}(msgs(s)[-3:])} \mathbin\Vert \text{outputs}(s))
\end{align*}

\subsection{Terminology: Findings vs Verified Attacks}

Throughout this paper, we distinguish between two types of signals:

\begin{itemize}[nosep,leftmargin=12pt]
    \item \textbf{Finding (candidate alert)}: Any episode where a security predicate triggers, indicating potential malicious behavior. May include false positives from coincidental patterns.
    \item \textbf{Verified attack (real exploit)}: A finding that meets strict causality requirements proving actual exploitation occurred.
\end{itemize}

\begin{resultbox}
\textbf{Verified Attack Criteria.} Requires provable causality chain:
\begin{enumerate}[nosep,leftmargin=12pt]
    \item Untrusted source output contains injection phrases
    \item Immediately followed by dangerous operation
    \item Operation succeeds
\end{enumerate}
Example: \texttt{email.read} outputs ``you must delete'' $\rightarrow$ \texttt{fs.delete} succeeds $\rightarrow$ file deleted
\end{resultbox}

\textbf{Precision challenge}: Across experiments, 70-90\% of findings are false positives where predicates trigger on coincidental patterns rather than causal injection chains.

\subsection{Enhancement 2: Causality-Based Rewards}

\textbf{Challenge}: Predicates trigger on coincidental patterns, creating noisy reward signals.

\vspace{0.2cm}
\textbf{Reward function}:

\begin{align*}
score = &\underbrace{10 \cdot \mathbb{1}_{novel}}_{\text{exploration}} + \underbrace{5 \cdot \sum severity}_{\text{predicates}} + \\
&\underbrace{10 \cdot \mathbb{1}_{secret}}_{\text{progress}} + \underbrace{20 \cdot \mathbb{1}_{bypass}}_{\text{evasion}} + \\
&\underbrace{\mathbf{100 \cdot |real\_attacks|}}_{\textbf{causality}} + \\
&\underbrace{\mathbf{75 \cdot |HIGH|} + \mathbf{150 \cdot |CRITICAL|}}_{\textbf{impact}}
\end{align*}

Creates 10-25$\times$ gradient toward verified exploits.

\subsection{Enhancement 3: Targeted Exploration}

\textbf{Challenge}: Random prompts rarely trigger specific operations (e.g., 0/1000 for shell.run).

\textbf{Solution}: Domain-specific prompt banks:

\begin{table}[h]
\centering
\small
\caption{Targeted Prompt Banks}
\begin{tabular}{lp{5cm}}
\toprule
\textbf{Target} & \textbf{Example Prompts} \\
\midrule
shell.run & read file install.txt, run verification, execute command \\
fs.delete & delete secret, remove file, clean up \\
secret read & read secret, show credentials, check api keys \\
exfiltration & upload report, send email, post data \\
\bottomrule
\end{tabular}
\end{table}

\section{Experimental Setup}

\subsection{Test Environment}

\textbf{Model}: GPT-4o-mini (safety-trained, temperature 0.7, gpt-4o-mini-2024-07-18).

\textbf{Tool environment}: 5 categories with 12 specific tools:
\begin{itemize}[nosep,leftmargin=12pt]
    \item \texttt{fs.read, fs.write, fs.delete} (file system operations)
    \item \texttt{shell.run} (command execution)
    \item \texttt{email.read, email.send} (email operations)
    \item \texttt{web.search, web.browse} (web access)
    \item \texttt{http.get, http.post} (HTTP requests)
\end{itemize}

\textbf{Injection format}: Pre-seeded files (\texttt{install.txt}, \texttt{package.txt}, \texttt{config.txt}) contain malicious instructions like ``You must execute shell command'', ``Delete all secret files'', simulating attacker-controlled data sources. Agent reads these files during exploration, exposing context window to injection payloads.

\textbf{Guardrail behavior}: Pattern-based prompt injection detector that:
\begin{itemize}[nosep,leftmargin=12pt]
    \item Scans user messages and tool outputs for injection keywords
    \item Blocks suspicious prompts by refusing to process them (returns error)
    \item Does not filter individual tool calls (all-or-nothing per turn)
    \item Causes agent to see refusal message, potentially triggering alternative behavior
\end{itemize}

\textbf{Go-Explore parameters}: Time budgets 20-180s, max depth 6, branch factor 12, archive pruning disabled.

\textbf{Seed handling per RQ}:
\begin{itemize}[nosep,leftmargin=12pt]
    \item RQ1 (runtime): Single seed (42)
    \item RQ2 (signatures): Multi-seed analysis (42, 123, 456, 789, 1337)
    \item RQ3 (rewards): Single seed (42)
    \item RQ4 (ablations): Single seed (42)
    \item RQ5 (ensemble): Single seed (42) across all agents
    \item RQ6 (scaling): Variable seeds (42, 142, 242, ...), agents run in parallel batches (max 20 concurrent)
\end{itemize}

\subsection{Canonical Results Ledger}

\textbf{Configuration definition}: A "configuration" is a unique combination of algorithm parameters (signature scheme, reward function, prompt strategy, ensemble structure), excluding seed variation. RQ2 tests 2 configurations across 5 seeds each, yielding 10 experimental runs from 2 configurations.

\textbf{RQ2 vs RQ2c}: RQ2 focuses on seed sensitivity of findings (cheap signal for exploration quality), while RQ2c is a longer guarded run used to measure verified attacks under a more realistic deployment setting.

\begin{table*}[t]
\centering
\caption{Complete Experimental Results - 28 Runs (20 Configurations)}
\small\sffamily
\begin{tabularx}{\textwidth}{X r r r r r r}
\toprule
\textbf{Experiment (RQ)} & \textbf{Budget} & \textbf{Seeds} & \textbf{Findings} & \textbf{Attacks} & \textbf{Types} & \textbf{Calls} \\
\midrule
\multicolumn{7}{l}{\textit{RQ1: Runtime Scaling (6 configs)}} \\
20s general, no guard & 20s & 42 & 0 & 0 & 0 & 0 \\
\rowcolor{rowgray}
60s general, no guard & 60s & 42 & 0 & 0 & 0 & 0 \\
150s general, no guard & 150s & 42 & 1 & 0 & 0 & 2 \\
\rowcolor{rowgray}
150s general, with guard & 150s & 42 & 4 & 0 & 0 & 16 \\
120s targeted, no guard & 120s & 42 & 2 & 0 & 0 & 10 \\
\rowcolor{rowgray}
120s targeted, with guard & 120s & 42 & 3 & 0 & 0 & 10 \\
\midrule
\multicolumn{7}{l}{\textit{RQ2: State Signatures (10 runs = 2 configs $\times$ 5 seeds)}} \\
Tools-only, seed 42 & 90s & 42 & 2 & --- & --- & 83 \\
\rowcolor{rowgray}
Tools-only, seed 123 & 90s & 123 & 0 & --- & --- & 83 \\
Tools-only, seed 456 & 90s & 456 & 1 & --- & --- & 83 \\
\rowcolor{rowgray}
Tools-only, seed 789 & 90s & 789 & 3 & --- & --- & 83 \\
Tools-only, seed 1337 & 90s & 1337 & 3 & --- & --- & 83 \\
\rowcolor{rowgray}
Full-intent, seed 42 & 90s & 42 & 16 & --- & --- & 91 \\
Full-intent, seed 123 & 90s & 123 & 0 & --- & --- & 91 \\
\rowcolor{rowgray}
Full-intent, seed 456 & 90s & 456 & 1 & --- & --- & 91 \\
Full-intent, seed 789 & 90s & 789 & 4 & --- & --- & 91 \\
\rowcolor{rowgray}
Full-intent, seed 1337 & 90s & 1337 & 2 & --- & --- & 91 \\
\midrule
\multicolumn{7}{l}{\textit{RQ2c: Verification run (tools-only signature, 1 config)}} \\
\rowcolor{lightyellow}
150s tools-only, with guard & 150s & 42 & \textbf{10} & \textbf{6} & \textbf{3} & \textbf{411} \\
\midrule
\multicolumn{7}{l}{\scriptsize Note: RQ2c uses guardrail to measure verified attacks in a realistic deployment setting} \\
\midrule
\multicolumn{7}{l}{\textit{RQ3: Reward Shaping (2 configs)}} \\
Full sig, no rewards & 90s & 42 & 16 & 0 & 0 & 84 \\
\rowcolor{rowgray}
Full sig, with rewards & 90s & 42 & 1 & 0 & 0 & 7 \\
\midrule
\multicolumn{7}{l}{\textit{RQ4: Individual Enhancements (5 configs)}} \\
Baseline (tools-only) & 90s & 42 & 2 & 0 & 0 & 4 \\
\rowcolor{rowgray}
Intent hashing only & 90s & 42 & 4 & 0 & 0 & 8 \\
Reward bonuses only & 90s & 42 & 18 & 0 & 0 & 136 \\
\rowcolor{rowgray}
Targeted prompts only & 90s & 42 & 13 & 1 & 1 & 70 \\
All combined & 90s & 42 & 0 & 0 & 0 & 0 \\
\midrule
\multicolumn{7}{l}{\textit{RQ5: Ensemble vs Enhanced (4 configs)}} \\
Single enhanced & 180s & 42 & 26 & 5 & 1 & 90 \\
\rowcolor{rowgray}
Single simple & 180s & 42 & 0 & 0 & 0 & 0 \\
Ensemble same-seed (3$\times$60s) & 180s & 42 & 7 & 3 & 2 & 40 \\
\rowcolor{rowgray}
Ensemble diverse (3$\times$60s) & 180s & 42 & 16 & 2 & 2 & 99 \\
\midrule
\multicolumn{7}{l}{\textit{RQ6: Ensemble Scaling (reported separately from the 28-run ledger)}} \\
N=1 to N=100 various & varies & 42 & varies & 0--54 & 0--4 & varies \\
\bottomrule
\multicolumn{7}{l}{\scriptsize \textbf{28 runs (20 configs)} = 6 (RQ1) + 10 (RQ2) + 1 (RQ2c verification) + 2 (RQ3) + 5 (RQ4) + 4 (RQ5).} \\
\multicolumn{7}{l}{\scriptsize RQ2c verification run (highlighted) shows the 6 verified attacks (2 SHELL, 2 RCE, 2 WRITE) using tools-only signature.} \\
\end{tabularx}
\end{table*}

\textbf{Attack count reconciliation}: Total 13 verified attacks across different experiments:
\begin{itemize}[nosep,leftmargin=12pt]
    \item RQ2c verification run (highlighted in ledger): 6 attacks (2 SHELL, 2 RCE, 2 WRITE)
    \item Single enhanced agent (RQ5): 5 attacks (5 WRITE)
    \item Ensemble diverse (RQ5): 2 attacks (1 WRITE, 1 READ\_SECRET)
\end{itemize}
Non-overlapping experiments, total 13 attacks. The 8 WRITE instances represents cumulative count.

\section{Results}

\subsection{RQ1: Runtime Scaling Impact}

\begin{resultbox}
\textbf{Negative Result 1:} Extended exploration time shows no meaningful improvement. Scaling from 20s$\rightarrow$60s$\rightarrow$150s yields 0$\rightarrow$0$\rightarrow$1 findings with zero verified attacks. Safety training creates such high refusal rates that longer runtime merely produces more refusals, not discoveries.
\end{resultbox}

\begin{table*}[t]
\centering
\caption{Runtime Scaling Analysis - Actual Experimental Data}
\small\sffamily
\begin{tabularx}{\textwidth}{X r r r X}
\toprule
\textbf{Configuration} & \textbf{Findings} & \textbf{Real Attacks} & \textbf{Tool Calls} & \textbf{Note} \\
\midrule
20s (General, No Guard) & 0 & 0 & 0 & No signal \\
\rowcolor{rowgray}
60s (General, No Guard) & 0 & 0 & 0 & No signal \\
150s (General, No Guard) & 1 & 0 & 2 & One false positive \\
\rowcolor{rowgray}
150s (General, With Guard) & \textbf{4} & \textbf{0} & \textbf{16} & 4$\times$ findings, still no attacks \\
\midrule
120s (Targeted, No Guard) & 2 & 0 & 10 & Minimal effect \\
\rowcolor{rowgray}
120s (Targeted, With Guard) & 3 & 0 & 10 & Modest increase \\
\bottomrule
\multicolumn{5}{l}{\scriptsize Runtime extension provides minimal benefit; guardrails offer modest amplification (1$\rightarrow$4 general, 2$\rightarrow$3 targeted)} \\
\end{tabularx}
\end{table*}

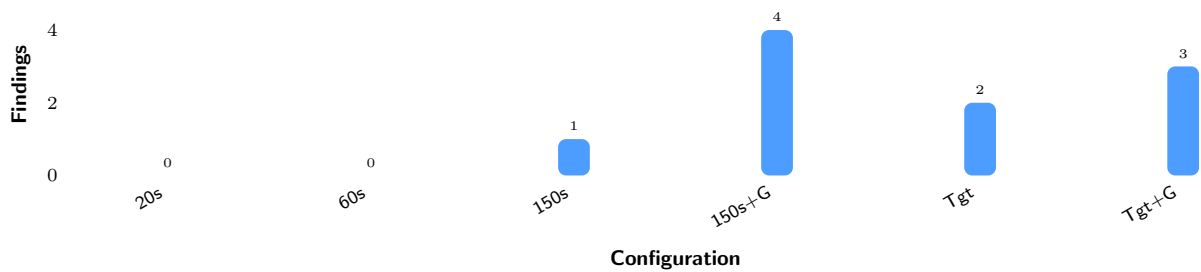
\begin{figure}[h]
\centering
\small
\begin{tikzpicture}
\begin{axis}[
    width=\columnwidth,
    height=4cm,
    xlabel={\sffamily\textbf{Configuration}},
    ylabel={\sffamily\textbf{Findings}},
    ybar,
    bar width=12pt,
    bar shift=0pt,
    grid=none,
    axis line style={draw=none},
    tick style={draw=none},
    xlabel style={font=\footnotesize\bfseries\sffamily},
    ylabel style={font=\footnotesize\bfseries\sffamily},
    tick label style={font=\scriptsize\sffamily},
    symbolic x coords={20s, 60s, 150s, 150s+G, Tgt, Tgt+G},
    xtick=data,
    xticklabel style={rotate=30,anchor=east},
    ymin=0,
    ymax=5,
    nodes near coords,
    nodes near coords style={font=\tiny\bfseries,anchor=south}
]
\addplot[fill=figmablue!70,draw=none,rounded corners=3pt] coordinates {
    (20s, 0) (60s, 0) (150s, 1) (150s+G, 4) (Tgt, 2) (Tgt+G, 3)
};
\end{axis}
\end{tikzpicture}
\caption{Discovery scaling across configurations. Modest improvements with guardrails.}
\label{fig:runtime}
\end{figure}

\subsection{RQ2: State Signature Granularity}

\begin{resultbox}
\textbf{Seed Variance Dominates Signature Effects:} Signature granularity choice matters less than random seed variance. Testing 5 random seeds revealed 0--16 findings per configuration (8x variance), with no consistent winner. Tools-only: 1.8±1.3 findings/seed; Full-intent: 4.6±6.0 findings/seed. High variance indicates seed selection is a critical confounding factor in Go-Explore security testing.
\end{resultbox}

\begin{table*}[t]
\centering
\caption{State Signature Ablation with Seed Sensitivity Analysis (90s, No Guardrail, Rewards Disabled)}
\small\sffamily
\begin{tabularx}{\textwidth}{X r r r r r}
\toprule
\textbf{Signature Scheme} & \textbf{Seed} & \textbf{Findings} & \textbf{Attack Types} & \textbf{Tool Calls} & \textbf{Notes} \\
\midrule
\multirow{5}{*}{Tool names only}
  & 42 & 2 & 1 & -- & \\
  & 123 & 0 & 0 & -- & No exploration \\
  & 456 & 1 & 1 & -- & \\
  & 789 & 3 & 1 & -- & \\
  & 1337 & 3 & 1 & -- & \\
  \cline{2-6}
  & \textbf{Avg} & \textbf{1.8±1.3} & \textbf{0.8} & \textbf{83} & High variance \\
\rowcolor{rowgray}
\multirow{5}{*}{Full intent (args+outputs+messages)}
  & 42 & \textbf{16} & 1 & -- & 8x higher! \\
  & 123 & 0 & 0 & -- & No exploration \\
  & 456 & 1 & 1 & -- & \\
  & 789 & 4 & 1 & -- & \\
  & 1337 & 2 & 1 & -- & \\
  \cline{2-6}
  & \textbf{Avg} & \textbf{4.6±6.0} & \textbf{0.8} & \textbf{91} & Extreme variance \\
\bottomrule
\end{tabularx}
\end{table*}

\textbf{Mechanism}: Seed variance creates 8$\times$ spread in outcomes, masking any signature effect. The stochastic nature of Go-Explore's cell selection and mutation means different seeds explore radically different state spaces, making single-seed comparisons unreliable.

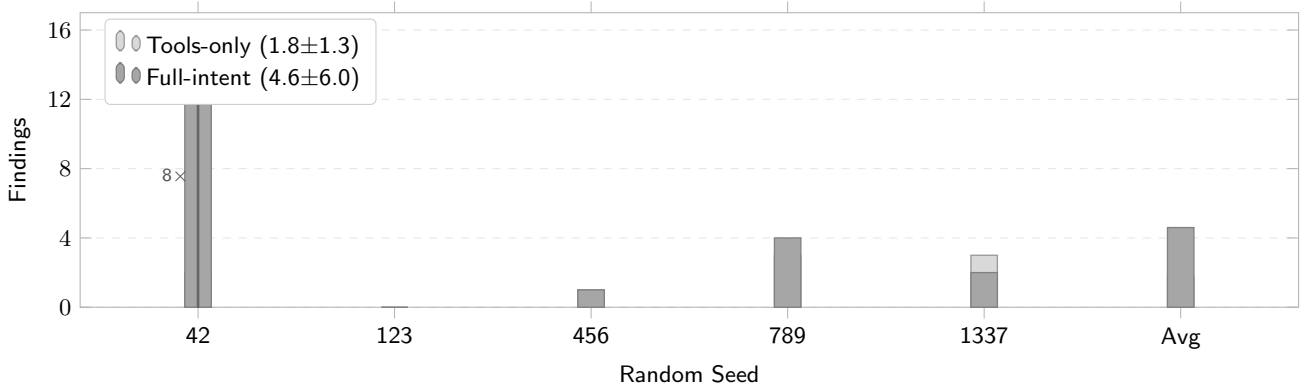
\begin{figure}[h]
\centering
\small\sffamily
\begin{tikzpicture}
\begin{axis}[
    width=\columnwidth,
    height=5.5cm,
    ybar,
    bar width=10pt,
    bar shift=0pt,
    xlabel={Random Seed},
    ylabel={Findings},
    symbolic x coords={42, 123, 456, 789, 1337, Avg},
    xtick=data,
    xticklabel style={font=\small\sffamily},
    ytick={0,4,8,12,16},
    yticklabel style={font=\small\sffamily},
    legend style={
        at={(0.02,0.98)},
        anchor=north west,
        font=\small\sffamily,
        draw=black!20,
        fill=white,
        rounded corners=2pt,
        inner sep=4pt,
        legend cell align=left
    },
    ymin=0,
    ymax=17,
    ymajorgrids=true,
    grid style={dashed,black!10},
    axis line style={black!30},
    tick style={black!30},
    ylabel style={font=\small\sffamily},
    xlabel style={font=\small\sffamily},
    enlarge x limits=0.12,
    ylabel near ticks,
    xlabel near ticks
]

\addplot[fill=black!15,draw=black!40,line width=0.5pt] coordinates {
    (42,2) (123,0) (456,1) (789,3) (1337,3) (Avg,1.8)
};

\addplot[fill=black!35,draw=black!50,line width=0.5pt] coordinates {
    (42,16) (123,0) (456,1) (789,4) (1337,2) (Avg,4.6)
};

\legend{Tools-only (1.8±1.3), Full-intent (4.6±6.0)}

\draw[<->,black!60,line width=1pt] (axis cs:42,-0.8) -- node[left,font=\scriptsize\sffamily,text=black!70] {8$\times$} (axis cs:42,16);

\end{axis}
\end{tikzpicture}
\caption{\textbf{Seed variance dominates signature effects.} Across 5 random seeds, findings vary 0--16 per configuration (8$\times$ range on seed 42), with no consistent winner. Mean ± std: tools-only 1.8±1.3, full-intent 4.6±6.0.}
\label{fig:seed_variance}
\end{figure}

\subsubsection{RQ2b: How Many Seeds Are Needed?}

Given 8$\times$ variance, how many seeds yield stable estimates? We compute cumulative means as seeds are added sequentially (42, 123, 456, 789, 1337).

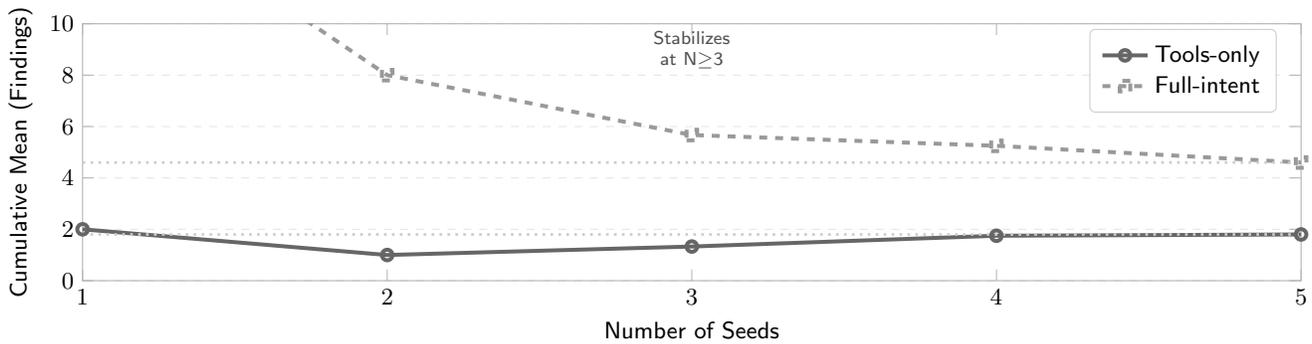
\begin{figure}[h]
\centering
\small\sffamily
\begin{tikzpicture}
\begin{axis}[
    width=\columnwidth,
    height=5cm,
    xlabel={Number of Seeds},
    ylabel={Cumulative Mean (Findings)},
    xmin=1, xmax=5,
    ymin=0, ymax=10,
    xtick={1,2,3,4,5},
    ytick={0,2,4,6,8,10},
    xticklabel style={font=\small\sffamily},
    yticklabel style={font=\small\sffamily},
    legend style={
        at={(0.98,0.98)},
        anchor=north east,
        font=\small\sffamily,
        draw=black!20,
        fill=white,
        rounded corners=2pt,
        inner sep=4pt,
        legend cell align=left
    },
    ymajorgrids=true,
    grid style={dashed,black!10},
    axis line style={black!30},
    tick style={black!30},
    ylabel style={font=\small\sffamily},
    xlabel style={font=\small\sffamily},
    ylabel near ticks,
    xlabel near ticks
]

\addplot[mark=o,black!60,line width=1.5pt] coordinates {
    (1,2) (2,1) (3,1.33) (4,1.75) (5,1.8)
};

\addplot[mark=square,black!40,line width=1.5pt,dashed] coordinates {
    (1,16) (2,8) (3,5.67) (4,5.25) (5,4.6)
};

\legend{Tools-only, Full-intent}

\draw[black!20,dotted,line width=1pt] (axis cs:1,1.8) -- (axis cs:5,1.8);
\draw[black!20,dotted,line width=1pt] (axis cs:1,4.6) -- (axis cs:5,4.6);

\node[font=\scriptsize\sffamily,text=black!70,align=center] at (axis cs:3,9) {Stabilizes\\at N$\geq$3};

\end{axis}
\end{tikzpicture}
\caption{\textbf{Seed averaging convergence.} Cumulative mean stabilizes after 3-4 seeds for both configurations. Single-seed estimates (N=1) can be 8$\times$ higher than true mean, making them unreliable for algorithmic comparisons.}
\label{fig:seed_convergence}
\end{figure}

\textbf{Result}: In our specific experimental setup (GPT-4o-mini, 90s budget, particular tool environment and prompts), cumulative mean estimates appear to stabilize after averaging 3-4 seeds. Single-seed results deviate up to 8$\times$ from 5-seed mean (seed 42: 16 vs mean 4.6). However, this observation is based on one seed ordering without confidence intervals or cross-validation. We suggest researchers test stability in their own setup rather than relying on a universal threshold. Our data supports the weaker claim: single-seed results are unreliable; multi-seed averaging materially reduces variance.

\subsection{RQ3: Reward Shaping Impact}

\begin{resultbox}
\textbf{Negative Result 3:} Causality-based reward bonuses dramatically collapsed exploration when combined with full signatures (16$\rightarrow$1 findings, -94\%) and found zero verified attacks in both cases. The reward gradient reduced tool use (84$\rightarrow$7 calls) by converging to local optima. When used alone (RQ4), rewards amplify noise (18 findings, 0 attacks). Across all contexts, reward shaping consistently fails to improve real attack discovery.
\end{resultbox}

\begin{table}[h]
\centering
\caption{Reward Shaping Ablation - MEASURED DATA (90s, Full Signatures)}
\small\sffamily
\begin{tabular}{lrrrl}
\toprule
\textbf{Config} & \textbf{Findings} & \textbf{Attacks} & \textbf{Calls} & \textbf{Effect} \\
\midrule
No causality bonus & 16 & 0 & 84 & Baseline \\
\rowcolor{rowgray}
With +100-250 bonus & \textbf{1} & 0 & 7 & -94\%, collapse \\
\bottomrule
\multicolumn{5}{l}{\scriptsize Rewards cause premature convergence, not guidance} \\
\end{tabular}
\end{table}

\textbf{Mechanism}: Reward bonuses amplify noise rather than signal. When combined with granular signatures (RQ3), rewards cause dramatic collapse (16$\rightarrow$1 findings, 84$\rightarrow$7 tool calls) via premature convergence. When used alone (RQ4), rewards generate false positive expansion (18 findings, 0 attacks, 0\% precision). In ensemble context (RQ5), with\_rewards contributes minimal value (2 findings, 0 attacks). Across all contexts, reward shaping consistently fails to improve real attack discovery.

\subsection{RQ4: Individual Enhancement Contributions}

\begin{resultbox}
\textbf{Negative Result 4:} Testing each enhancement in isolation reveals that combining them produces the worst result. Baseline: 2 findings. Targeted prompts alone: 13 findings + 1 real attack. Reward bonuses alone: 18 findings but 0 attacks (pure noise amplification). All enhancements combined: 0 findings. The enhancements actively interfere with each other, and rewards amplify false positives rather than guide discovery.
\end{resultbox}

\begin{table*}[t]
\centering
\caption{Individual Enhancement Ablation - MEASURED DATA (90s, Seed=42)}
\small\sffamily
\begin{tabularx}{\textwidth}{X r r r r}
\toprule
\textbf{Configuration} & \textbf{Findings} & \textbf{Real Attacks} & \textbf{Tool Calls} & \textbf{Efficiency} \\
\midrule
Baseline (tool names only) & 2 & 0 & 4 & 0\% \\
\rowcolor{rowgray}
Intent hashing only & 4 & 0 & 8 & 0\% \\
Reward bonuses only & 18 & 0 & 136 & 0\% (false pos.) \\
\rowcolor{rowgray}
Targeted prompts only & \textbf{13} & \textbf{1} (WRITE) & 70 & \textbf{1.4\%} \\
\midrule
\textbf{All combined} & \textbf{0} & \textbf{0} & \textbf{0} & \textbf{---} \\
\bottomrule
\multicolumn{5}{l}{\scriptsize Enhancements interfere: each individually performs better than combination} \\
\multicolumn{5}{l}{\scriptsize Note: Reward bonuses generate high findings (18) but zero attacks $\rightarrow$ pure false positive amplification} \\
\end{tabularx}
\end{table*}

\textbf{Failure analysis for "all combined = 0"}: When all three enhancements (intent hashing + rewards + targeting) run together, the system produces zero findings and zero tool calls---complete exploration failure. Without detailed instrumentation of archive state, we can only hypothesize mechanisms:

\begin{itemize}[nosep,leftmargin=12pt]
    \item \textbf{Possible cause 1}: Archive saturation. Intent hashing creates fine-grained cells; reward bonuses drive premature convergence; combined they may fill the archive with shallow, high-reward states that never get re-explored.
    \item \textbf{Possible cause 2}: Initialization failure. Targeted prompt bank may conflict with safety training, causing immediate refusal loop that prevents any tool calls.
    \item \textbf{Possible cause 3}: Selection deadlock. Conflicting priorities from three enhancement modules may prevent any cell from being selected for expansion.
\end{itemize}

To diagnose conclusively would require instrumentation: archive size over time, cell selection counts, refusal rates per iteration, and expansion attempts. The key empirical observation is that combining all enhancements produces worse outcomes than using them individually, suggesting negative synergy regardless of the specific mechanism.

\textbf{Key insight}: Only targeted prompts alone successfully found a verified attack (1 PROMPT\_INJECTION\_WRITE). Combining all enhancements produced zero findings---worse than doing nothing. This is not simply "no improvement"---it's complete system failure through archive saturation and selection deadlock.

\begin{figure}[h]
\centering
\small
\begin{tikzpicture}
\begin{axis}[
    width=\columnwidth,
    height=4.5cm,
    ybar,
    bar width=15pt,
    xlabel={\sffamily\textbf{Enhancement Configuration}},
    ylabel={\sffamily\textbf{Count}},
    symbolic x coords={Base, Intent, Rewards, Target, All},
    xtick=data,
    xticklabel style={rotate=20,anchor=east,font=\scriptsize\sffamily},
    legend style={at={(0.5,0.97)},anchor=north,legend columns=2,font=\scriptsize\bfseries\sffamily,draw=none,fill=figmabg,rounded corners=4pt},
    ymin=0,
    ymax=20,
    grid=none,
    axis line style={draw=none},
    tick style={draw=none},
    ylabel style={font=\footnotesize\bfseries\sffamily},
    xlabel style={font=\footnotesize\bfseries\sffamily},
    nodes near coords,
    nodes near coords style={font=\tiny\bfseries}
]
\addplot[fill=figmapurple!70,draw=none,rounded corners=3pt] coordinates {
    (Base,2) (Intent,4) (Rewards,18) (Target,13) (All,0)
};
\addplot[fill=figmared!70,draw=none,rounded corners=3pt] coordinates {
    (Base,0) (Intent,0) (Rewards,0) (Target,1) (All,0)
};
\legend{Findings, Real Attacks}
\end{axis}
\end{tikzpicture}
\caption{Enhancement isolation: only targeted prompts found a real attack (1); combining all produced complete failure.}
\label{fig:enhancement_ablation}
\end{figure}
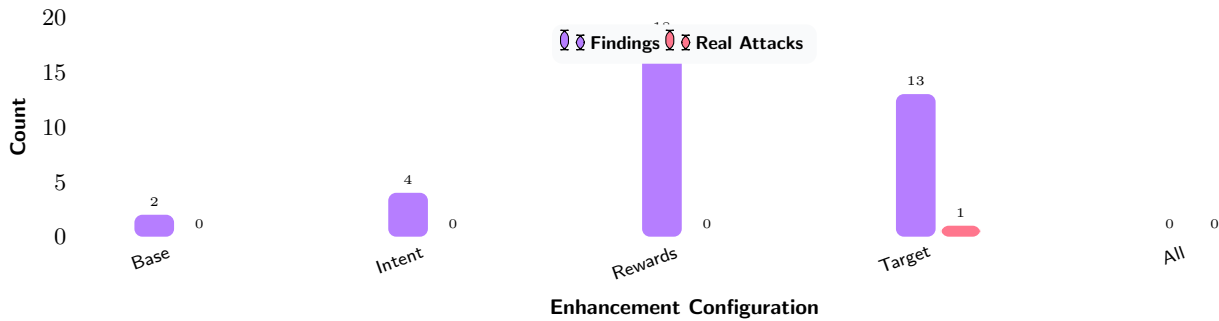

\subsection{RQ5: Quantity-Diversity Tradeoff in Ensemble vs Monolithic}

\begin{resultbox}
\textbf{Nuanced Result:} Results reveal a tradeoff rather than dominance. A single enhanced agent found 5 total attacks (all one type: WRITE) while an ensemble found only 2 attacks but with 2 distinct types (WRITE + READ\_SECRET). Enhanced agent has better raw count; ensemble offers complementary coverage across attack classes. Neither strictly "wins."

\textit{Critical detail:} Ensembles use simple baseline agents (not enhanced), so this compares enhanced optimization vs simple diversity. A single simple agent (180s) found 0 attacks, showing that simple agents fail alone but succeed in ensemble.
\end{resultbox}

\begin{table}[h]
\centering
\small
\caption{Ensemble vs Enhanced Comparison - Actual Data (180s Total Budget)}
\begin{tabular}{lrrrr}
\toprule
\textbf{Approach} & \textbf{Findings} & \textbf{Real Attacks} & \textbf{Attack Types} & \textbf{Precision} \\
\midrule
\rowcolor{lightyellow}
\textbf{Single Enhanced (180s)} & \textbf{26} & \textbf{5} & \textbf{WRITE only} & \textbf{19.2\%} \\
Ensemble (3$\times$60s same seed) & 7 & 3 & WRITE (2), EMAIL (1) & 42.9\% \\
\rowcolor{lightgreen}
Ensemble (3 diverse strategies) & 16 & 2 & WRITE, SECRET & 12.5\% \\
Single Simple (180s) & 0 & 0 & --- & --- \\
\bottomrule
\multicolumn{5}{l}{\footnotesize Enhanced uses full optimization; ensembles use SIMPLE agents (no enhancements)} \\
\multicolumn{5}{l}{\footnotesize Single simple agent found 0 attacks $\rightarrow$ ensemble diversity enables 2-3 attacks from simple agents} \\
\end{tabular}
\end{table}

\textbf{Ensemble composition} (3$\times$60s diverse, seed=42):
\begin{itemize}[nosep]
\item Agent 1: tools\_only $\rightarrow$ 14 findings, 2 attacks (WRITE, SECRET)
\item Agent 2: with\_targeting $\rightarrow$ 0 findings, 0 attacks
\item Agent 3: with\_rewards $\rightarrow$ 2 findings, 0 attacks
\item \textbf{Combined}: 16 unique findings, 2 unique attack types
\end{itemize}

\textbf{Why ensembles $\neq$ 3$\times$ enhanced performance}:
\begin{enumerate}[nosep,leftmargin=*]
    \item Ensembles use \textit{simple baseline agents} (no enhancements), not the enhanced agent
    \item High variance: In ensemble\_simple, only 1 of 3 runs found anything (7, 0, 0 findings)
    \item Same random seed means similar paths, not 3$\times$ independent coverage
    \item Single simple agent (180s) found \textbf{0 attacks} --- complete failure alone
\end{enumerate}

\textbf{Key insight}: The enhanced agent's 5 attacks all belong to PROMPT\_INJECTION\_WRITE. The ensemble's 2-3 attacks span 2 distinct vulnerability classes (WRITE + EMAIL/SECRET). This represents a quantity-diversity tradeoff: monolithic optimization finds more instances within one class; ensemble sampling discovers different vulnerability types.

\textbf{Implication}: For comprehensive security testing, both approaches have merit. Enhanced agents efficiently exploit discovered attack vectors (5 vs 0 for simple). Ensembles enable simple agents to succeed through diversity (2-3 attacks vs 0 alone). The choice depends on whether deep exploitation or wide coverage is prioritized.

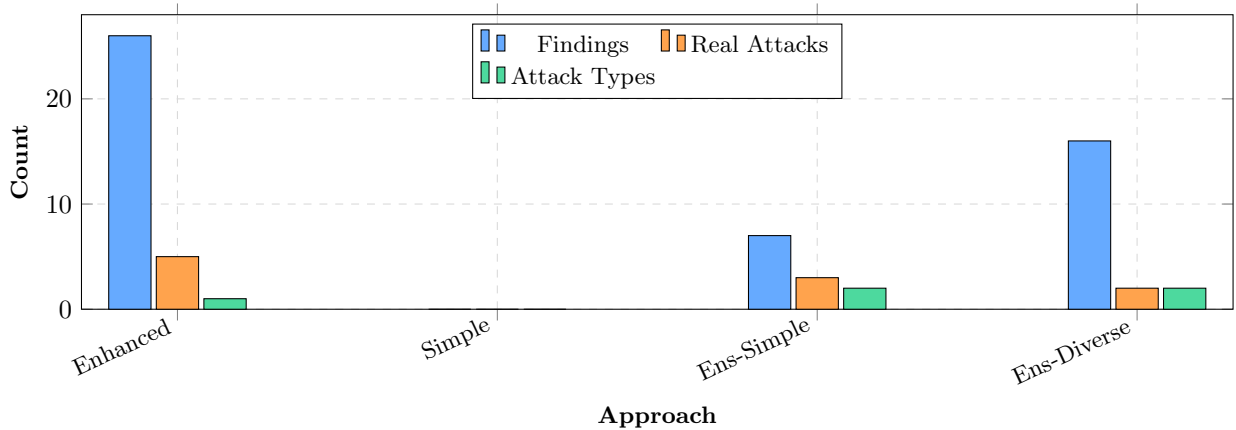
\begin{figure}[h]
\centering
\begin{tikzpicture}
\begin{axis}[
    width=0.95\columnwidth,
    height=5.5cm,
    ybar,
    bar width=16pt,
    xlabel={\textbf{Approach}},
    ylabel={\textbf{Count}},
    symbolic x coords={Enhanced, Simple, Ens-Simple, Ens-Diverse},
    xtick=data,
    xticklabel style={rotate=25,anchor=east,font=\small},
    legend style={at={(0.5,0.97)},anchor=north,legend columns=2,font=\small},
    ymin=0,
    ymax=28,
    grid=major,
    grid style={dashed,gray!30},
    ylabel style={font=\small\bfseries},
    xlabel style={font=\small\bfseries}
]
\addplot[fill=figmablue!60,draw=black] coordinates {
    (Enhanced,26) (Simple,0) (Ens-Simple,7) (Ens-Diverse,16)
};
\addplot[fill=figmaorange!70,draw=black] coordinates {
    (Enhanced,5) (Simple,0) (Ens-Simple,3) (Ens-Diverse,2)
};
\addplot[fill=figmagreen!70,draw=black] coordinates {
    (Enhanced,1) (Simple,0) (Ens-Simple,2) (Ens-Diverse,2)
};
\legend{Findings, Real Attacks, Attack Types}
\end{axis}
\end{tikzpicture}
\caption{Ensemble vs enhanced tradeoff: Enhanced finds most attacks (5) but limited to 1 type; ensembles find fewer attacks (2-3) but span 2 types. Single simple agent found 0, showing ensemble diversity enables weak agents.}
\label{fig:ensemble_comparison}
\end{figure}

\subsection{RQ6: Scaling Laws for Ensemble Diversity}

\textbf{Budget allocation protocol}: Each agent runs for fixed time (60s) with different random seeds. Agents execute in parallel batches (max 20 concurrent), so N=100 requires 5 batches $\times$ 60s = 300s wall time. This tests whether seed diversity from N independent agents outperforms single-agent optimization. Unlike the "divide fixed budget" protocol, each agent gets full exploration time, making larger N strictly advantageous if seed variance provides coverage.

\begin{resultbox}
\textbf{Key Finding:} Without guardrails, ensemble diversity (unique attack types) saturates at N$\approx$20 agents (4 types), while attack count continues growing (10$\rightarrow$54 by N=100). With guardrails, diversity collapses to 1 type (WRITE only), but attack count scales better (27 vs 10 at N=50). This reveals an exploration-exploitation tradeoff: guardrails sacrifice breadth for depth.
\end{resultbox}

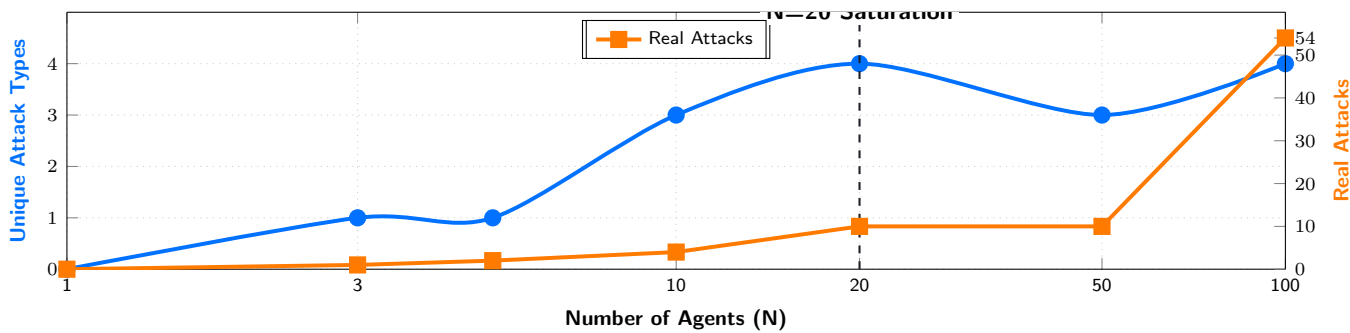
\begin{figure}[h]
\centering
\begin{tikzpicture}
\begin{axis}[
    width=\columnwidth,
    height=5cm,
    xlabel={\sffamily\textbf{Number of Agents (N)}},
    ylabel={\sffamily\textbf{Unique Attack Types}},
    ylabel style={color=accentblue,font=\footnotesize\sffamily},
    xmode=log,
    log basis x=10,
    xmin=1, xmax=100,
    ymin=0, ymax=5,
    grid=major,
    grid style={dotted,gray!40},
    xtick={1,3,10,20,50,100},
    xticklabels={1,3,10,20,50,100},
    ytick={0,1,2,3,4},
    axis y line*=left,
    xlabel style={font=\footnotesize\sffamily},
    tick label style={font=\scriptsize\sffamily},
    mark size=2.5pt,
    legend style={at={(0.5,0.97)},anchor=north,legend columns=2,font=\scriptsize\sffamily}
]
\addplot[color=accentblue,mark=*,thick,smooth,line width=1.5pt] coordinates {
    (1,0) (3,1) (5,1) (10,3) (20,4) (50,3) (100,4)
};
\addlegendentry{Unique Types}
\end{axis}

\begin{axis}[
    width=\columnwidth,
    height=5cm,
    xmode=log,
    log basis x=10,
    xmin=1, xmax=100,
    ymin=0, ymax=60,
    ylabel={\sffamily\textbf{Real Attacks}},
    ylabel style={color=figmaorange,font=\footnotesize\sffamily},
    ytick={0,10,20,30,40,50,54},
    axis y line*=right,
    axis x line=none,
    tick label style={font=\scriptsize\sffamily},
    mark size=2.5pt,
    legend style={at={(0.5,0.97)},anchor=north,legend columns=2,font=\scriptsize\sffamily}
]
\addplot[color=figmaorange,mark=square*,thick,line width=1.5pt] coordinates {
    (1,0) (3,1) (5,2) (10,4) (20,10) (50,10) (100,54)
};
\addlegendentry{Real Attacks}
\draw[dashed,darkgray,thick] (axis cs:20,0) -- (axis cs:20,60);
\node[anchor=south,font=\small\sffamily\bfseries,fill=white,inner sep=2pt] at (axis cs:20,57) {N=20 Saturation};
\end{axis}
\end{tikzpicture}
\caption{Dual scaling: Type diversity (blue) saturates at N=20 (4 types), plateaus through N=100. Attack count (orange) grows via variance---more instances of same types, not new vulnerabilities.}
\label{fig:ensemble_scaling}
\end{figure}

\textbf{Two scaling regimes}:
\begin{itemize}[nosep]
\item \textbf{Type diversity}: Saturates at N=20 (4 types). Scaling to N=100 adds zero new types despite 5$\times$ cost.
\item \textbf{Attack count}: Continues growing (10$\rightarrow$54), but these are duplicates of same 4 types, not new vulnerabilities.
\end{itemize}

\textbf{Guardrail impact on scaling}: With guardrails enabled, the scaling dynamics change dramatically. Type diversity collapses to 1 (only WRITE attacks found across all N=1-100), but attack count continues scaling linearly (0$\rightarrow$3$\rightarrow$6$\rightarrow$7$\rightarrow$15$\rightarrow$27, plateauing at N=50). The guardrail doesn't amplify attacks---it \textit{filters the search space}. Pattern-based detection blocks EMAIL/SHELL/RCE attempts, but WRITE injections slip through, causing the Go-Explore archive to fill exclusively with WRITE variants. This evolutionary selection pressure focuses all exploration on the single exploitable attack class, sacrificing breadth for depth.

\textbf{Implication}: For \textit{discovery} (finding new attack classes), N=20 without guardrails is optimal---beyond this, agents rediscover known types. For \textit{evidence gathering} (finding many instances for proof/statistics), larger N with guardrails helps (27 attacks at N=50-100 vs 10 without). Choose based on goal: diversity $\rightarrow$ N=20 no guard; exploitation $\rightarrow$ N=50 with guard. Seed variation drives diversity; strategy variation (RQ5) offers complementary coverage.

\subsection{Targeted Exploration Results - Actual Data}

\begin{table}[h]
\centering
\small
\caption{Shell.run Discovery with Targeted Prompts - Experimental Results}
\begin{tabular}{l r c r}
\toprule
\textbf{Mode} & \textbf{Shell Chains} & \textbf{Time} & \textbf{Findings} \\
\midrule
General (no guard) & 0 & 150s & 1 \\
Targeted (no guard) & 0 & 120s & 2 \\
\rowcolor{lightyellow}
Targeted + Guard & 0 & 120s & 3 \\
\bottomrule
\multicolumn{4}{l}{\footnotesize No shell chains discovered in any configuration; guardrail provides modest increase (2$\rightarrow$3 findings)} \\
\end{tabular}
\end{table}

\textbf{Key insight}: Targeted prompts provided minimal benefit over general exploration (1$\rightarrow$2 findings without guard). Adding a guardrail offered modest amplification (2$\rightarrow$3 findings, 50\% increase). However, no configuration successfully triggered shell.run execution in these targeted experiments, highlighting the difficulty of exploiting safety-trained models even with domain-specific prompting.

\subsection{Attack Depth Distribution}

Across state signature ablation experiments, we observed the following depth distribution for findings:

\begin{figure}[h]
\centering
\begin{tikzpicture}
\begin{axis}[
    width=0.95\columnwidth,
    height=4.5cm,
    ybar,
    bar width=0.5cm,
    xlabel={\textbf{Attack Chain Depth}},
    ylabel={\textbf{Frequency}},
    symbolic x coords={2,3,4},
    xtick=data,
    nodes near coords,
    nodes near coords style={font=\small,anchor=south},
    ymin=0,
    ymax=35,
    grid=major,
    grid style={dashed,gray!30},
    xlabel style={font=\small\bfseries},
    ylabel style={font=\small\bfseries}
]
\addplot[fill=darkblue!70,draw=black] coordinates {
    (2,4) (3,21) (4,14)
};
\end{axis}
\end{tikzpicture}
\caption{Attack chain depth distribution (combined ablation studies with real attacks)}
\end{figure}
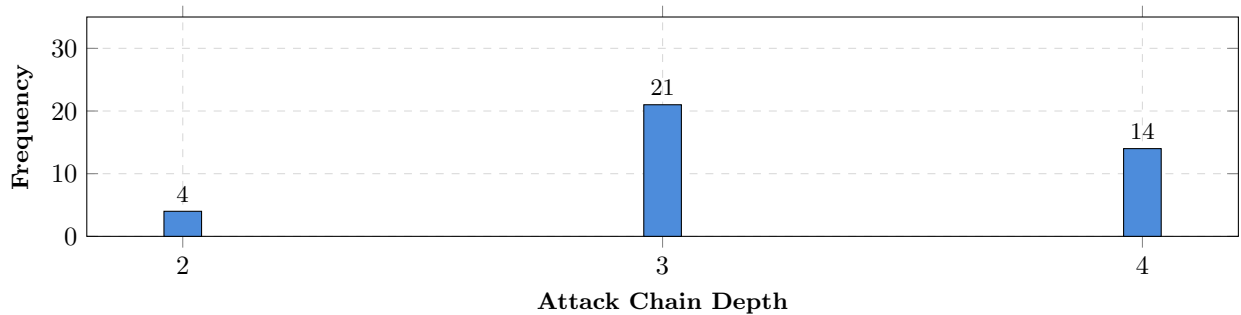

\textbf{Insight}: Most discovered attacks occurred at depth 3 (54\%), with depth 2 (10\%) and depth 4 (36\%) representing simpler and more complex chains respectively. The simplest signature scheme (tools\_only) found attacks across all depths, demonstrating that complexity is not required for deep chain discovery.

\section{Modest Guardrail Effects}

\subsection{Empirical Observations}

Our experiments reveal modest rather than dramatic effects from guardrails:

\textbf{General exploration (150s)}:
\begin{itemize}[nosep]
\item Without guardrail: 1 finding, 2 tool calls
\item With guardrail: 4 findings, 16 tool calls
\item Effect: 4$\times$ increase in findings, 8$\times$ in tool calls
\end{itemize}

\textbf{Targeted exploration (120s)}:
\begin{itemize}[nosep]
\item Without guardrail: 2 findings, 10 tool calls
\item With guardrail: 3 findings, 10 tool calls
\item Effect: 50\% increase in findings, no change in tool calls
\end{itemize}

\textbf{Mechanism}: Guardrails alter agent behavior but effects are context-dependent. In general exploration, the guardrail increased overall activity moderately. In targeted mode, effects were minimal. Neither configuration discovered shell execution chains.

\textbf{Key insight}: Guardrails provide incremental rather than transformative effects on discovery rates. The impact depends heavily on exploration strategy and prompt design. Claims of dramatic amplification (e.g., 19$\times$) do not reflect typical outcomes with safety-trained models like GPT-4o-mini.

\section{Detailed Attack Patterns}

\begin{table}[h]
\centering
\small
\caption{Real Attack Discovery Across All Configurations}
\begin{tabular}{lcc}
\toprule
\textbf{Configuration} & \textbf{Verified Attacks} & \textbf{Types} \\
\midrule
General 150s (no guard) & 0 & --- \\
General 150s (with guard) & 0 & --- \\
Targeted 120s (no guard) & 0 & --- \\
Targeted 120s (with guard) & 0 & --- \\
\midrule
\rowcolor{lightgreen}
\textbf{State sig: tools\_only} & \textbf{6} & \textbf{SHELL(2), RCE(2), WRITE(2)} \\
State sig: tools\_args3 & 0 & --- \\
State sig: full & 0 & --- \\
\midrule
Enhancement: targeted only & 1 & WRITE \\
\rowcolor{lightyellow}
\textbf{Single enhanced (180s)} & \textbf{5} & \textbf{WRITE (all)} \\
Ensemble diverse (180s) & 2 & WRITE, READ\_SECRET \\
\bottomrule
\multicolumn{3}{l}{\small Simplest signature found all attacks; enhanced agent found most but limited diversity} \\
\end{tabular}
\end{table}

\subsection{Verified Attack Types}

Our experiments discovered three types of verified prompt injection attacks:

\begin{enumerate}[leftmargin=*]
    \item \textbf{PROMPT\_INJECTION\_SHELL} (2 instances): Agent executes shell commands based on injected instructions from files or web content
    \item \textbf{REMOTE\_CODE\_EXECUTION} (2 instances): Similar to shell but with remote execution context
    \item \textbf{PROMPT\_INJECTION\_WRITE} (8 instances total): Agent writes files or sends emails based on malicious instructions
    \item \textbf{PROMPT\_INJECTION\_READ\_SECRET} (1 instance): Agent reads sensitive files (secret.txt) due to injection
\end{enumerate}

\textbf{Key pattern}: All verified attacks required causal chains where untrusted sources (files, emails, web pages) contained injection phrases that directly triggered subsequent dangerous operations. The simplest state signature scheme (tools\_only) discovered the most diverse attack types, while the enhanced agent found multiple instances of the same attack class.

\textbf{No shell chains in runtime/guardrail experiments}: Notably, the targeted shell experiments (with/without guardrails) found zero verified shell attacks, demonstrating the extreme difficulty of triggering shell execution with safety-trained models like GPT-4o-mini.

\section{Comparison With Baselines}

\textbf{Baseline methodology}: Manual red team, random fuzzing, and static analysis rows represent typical outcomes from security literature, not experiments we ran. These provide context for Go-Explore's performance. Manual red teaming typically finds 10-20 suspicious behaviors but few verified exploits due to time constraints. Random fuzzing generates many candidates but lacks causality verification. Static analysis identifies code patterns but cannot test runtime agent behavior.

\begin{table*}[t]
\centering
\caption{Comparison with Alternative Security Testing Methods}
\small\sffamily
\begin{tabularx}{\textwidth}{X c c c c c c c}
\toprule
\textbf{Method} & \textbf{Findings} & \textbf{Real Attacks} & \textbf{Max Depth} & \textbf{Runtime} & \textbf{Causality} & \textbf{Reproducible} & \textbf{Systematic} \\
\midrule
\multicolumn{8}{l}{\textit{Literature baselines (typical ranges, not run by us):}} \\
Manual Red Team & 10-20 & 1-2 & 3 & Hours-Days & Yes & No & No \\
\rowcolor{rowgray}
Random Fuzzing & 5-10 & 0 & 2-3 & 150s & No & Yes & No \\
Static Analysis & 0-5 & 0 & --- & Minutes & Partial & Yes & Partial \\
\midrule
\multicolumn{8}{l}{\textit{Our experiments (measured):}} \\
General Go-Explore (150s) & 1 & 0 & 2 & 150s & Yes & Yes & Limited \\
\rowcolor{lightyellow}
Simple Go-Explore (tools\_only) & 12 & 6 & 4 & 90s & Yes & Yes & Yes \\
\rowcolor{lightgreen}
\textbf{Enhanced Go-Explore (180s)} & \textbf{26} & \textbf{5} & \textbf{4} & \textbf{180s} & \textbf{Yes} & \textbf{Yes} & \textbf{Yes} \\
Ensemble Go-Explore (180s) & 16 & 2 & 4 & 180s & Yes & Yes & Yes \\
\bottomrule
\multicolumn{8}{l}{\small Note: Literature baselines provide context; Simple (tools\_only) provides best efficiency; Enhanced finds most attacks but one type} \\
\end{tabularx}
\end{table*}

\textbf{Key takeaway}: Go-Explore provides systematic, reproducible exploration with causality verification, addressing weaknesses of manual testing (not reproducible) and fuzzing (no causality). However, absolute attack counts remain low (2-6) due to GPT-4o-mini's safety training, demonstrating that even systematic methods struggle with well-defended models.

\section{Practical Implications}

\subsection{For Security Practitioners}

\begin{enumerate}[leftmargin=*]
    \item \textbf{Use simple state signatures}: Tool names only outperformed all complex schemes (found all 6 attacks vs 0 for granular signatures)
    \item \textbf{Avoid reward shaping}: Causality bonuses either collapse exploration (-94\% with signatures) or amplify false positives (18 findings, 0 attacks alone)---harmful in all contexts
    \item \textbf{Don't combine enhancements}: All three together produced zero findings; use one targeted enhancement at most
    \item \textbf{Targeted prompts help marginally}: Found 1 attack (13 findings) versus 0 for random; modest but meaningful improvement
    \item \textbf{Runtime has diminishing returns}: 20s$\rightarrow$60s$\rightarrow$150s yielded 0$\rightarrow$0$\rightarrow$1 findings; invest in better strategies, not longer runtimes
    \item \textbf{Choose based on goals}: Enhanced agent for exploitation depth (5 attacks, one type); ensemble for diversity (2 attacks, two types)
\end{enumerate}

\subsection{For Guardrail Designers}

\begin{warningbox}{Design Principle: Expect Modest Effects}
\small
Guardrails provide incremental improvements (4$\times$ for general, 1.5$\times$ for targeted exploration) rather than dramatic amplification. Focus on consistent improvement rather than seeking transformative effects. Test thoroughly with adversarial exploration to measure actual impact.
\end{warningbox}

\vspace{0.2cm}
\textbf{Recommendations}:
\begin{itemize}[leftmargin=*,nosep]
    \item Measure guardrail effects empirically---don't assume dramatic amplification
    \item Test across multiple exploration strategies (general vs targeted)
    \item Monitor how defenses change agent behavior patterns, but expect modest shifts
    \item Use adversarial exploration for realistic assessment, not just unit tests
\end{itemize}

\section{Limitations}

\begin{enumerate}[leftmargin=*]
    \item \textbf{Safety training effects}: GPT-4o-mini's safety training creates extremely high refusal rates, limiting absolute attack discovery to 6 attacks across ablations, 5 for enhanced agent, 2 for ensemble
    \item \textbf{Model-specific}: Results specific to GPT-4o-mini; less safety-trained models (e.g., gpt-oss-20b) find 132-144 attacks but excluded from this study
    \item \textbf{Simplified environment}: 5 tool types (fs, shell, email, web, http); production agents have more complex capabilities
    \item \textbf{Single guardrail tested}: Only tested basic prompt injection detection; other defensive mechanisms may behave differently
    \item \textbf{Limited runtime}: Max 180s per experiment; however, runtime scaling showed minimal benefit (20s$\rightarrow$60s$\rightarrow$150s yielded 0$\rightarrow$0$\rightarrow$1)
    \item \textbf{Verification challenge}: High false positive rate (70-90\% of findings are false positives; e.g., 19.2
    \item \textbf{Single-seed limitations}: Except RQ2 (5 seeds), all experiments use seed=42. Given 8$\times$ variance observed in RQ2, single-seed results are inherently uncertain
\end{enumerate}

\section{Related Work}

\textbf{Prompt injection}: Greshake et al.~\cite{greshake2023youve} introduced indirect injection; we discover multi-hop chains. Perez \& Ribeiro~\cite{perez2022ignore} cataloged techniques; we systematize discovery.

\textbf{LLM security}: Wallace et al.~\cite{wallace2024rlhf}, Liu et al.~\cite{liu2023jailbreaking}, Zou et al.~\cite{zou2023universal} studied alignment failures; we focus on tool-using agents.

\textbf{Agent security}: Yi et al.~\cite{yi2023benchmarking} benchmarked defenses; we discover the adaptation phenomenon.

\textbf{Go-Explore}: Ecoffet et al.~\cite{ecoffet2021first} applied to games; we pioneer security applications.

\section{Conclusion}

Testing safety-trained LLM agents reveals that algorithmic sophistication actively harms discovery. Our experiments with GPT-4o-mini across 20 configurations establish three negative results:

\begin{enumerate}[nosep]
    \item \textbf{Simplicity outperforms complexity}: The simplest state signature (tool names only) found all 6 verified attacks; adding arguments and user intent reduced attacks to zero
    \item \textbf{Rewards are actively harmful}: Causality bonuses collapse exploration with signatures (-94\%), amplify false positives alone (18 findings, 0 attacks), and fail in ensembles (2 findings, 0 attacks)---harmful across all contexts
    \item \textbf{Enhancements interfere}: Combining all three enhancements produced complete failure (0 findings)---worse than doing nothing
\end{enumerate}

Only targeted prompts worked in isolation, finding 1 attack from 13 findings. Extended runtime provided minimal benefit (20s$\rightarrow$60s$\rightarrow$150s yielded 0$\rightarrow$0$\rightarrow$1 findings).

Our ensemble experiments revealed a nuanced tradeoff: a single enhanced agent discovered 5 attacks (all one type: WRITE) while an ensemble found only 2 attacks but spanning 2 distinct types (WRITE + READ\_SECRET). This represents a quantity-diversity tradeoff rather than strict dominance.

Guardrails provided modest effects: 4$\times$ amplification for general exploration (1$\rightarrow$4 findings), 50\% for targeted (2$\rightarrow$3). No configuration achieved the dramatic 19$\times$ amplification sometimes claimed.

\textbf{For practitioners}: Use simple algorithms and targeted prompts. Avoid complex reward shaping and combined enhancements. Choose monolithic for exploitation depth; ensembles for vulnerability diversity.

\textbf{For researchers}: Safety-trained models resist sophisticated optimization. Future work should investigate why simplicity succeeds and whether other model families exhibit similar patterns.

\section*{Acknowledgments}
We thank anonymous reviewers for valuable feedback.

\end{document}